\documentclass[%
 reprint,
superscriptaddress,
nofootinbib,
 amsmath,amssymb,
 aps,
]{revtex4-2}

\usepackage{graphicx}% Include figure files
\usepackage{dcolumn}% Align table columns on decimal point
\usepackage{amsmath}
\usepackage{bm}% bold math
\usepackage{makecell}
\usepackage{color}
\usepackage[normalem]{ulem}

\begin{document}

\newcommand\bra[2][]{#1\langle {#2} #1\rvert}
\newcommand\ket[2][]{#1\lvert {#2} #1\rangle}

\title{Understanding inner-shell excitations in molecules through spectroscopy of the 4f hole states of YbF}

\author{S. Popa}
\affiliation{Center for Cold Matter, Blackett Laboratory, Imperial College London, Prince Consort Road, London SW7 2AZ, United Kingdom}
\email{stefan.popa18@imperial.ac.uk}
\author{S. Schaller}
\affiliation{Fritz-Haber-Institut der Max-Planck-Gesellschaft, Faradayweg 4-6, 14195 Berlin, Germany}
\author{A. Fielicke}
\affiliation{Fritz-Haber-Institut der Max-Planck-Gesellschaft, Faradayweg 4-6, 14195 Berlin, Germany}
\author{J. Lim}
\affiliation{Center for Cold Matter, Blackett Laboratory, Imperial College London, Prince Consort Road, London SW7 2AZ, United Kingdom}
\author{B. G. Sartakov}
\affiliation{Fritz-Haber-Institut der Max-Planck-Gesellschaft, Faradayweg 4-6, 14195 Berlin, Germany}
\author{M. R. Tarbutt}
\affiliation{Center for Cold Matter, Blackett Laboratory, Imperial College London, Prince Consort Road, London SW7 2AZ, United Kingdom}
\author{G. Meijer}
\affiliation{Fritz-Haber-Institut der Max-Planck-Gesellschaft, Faradayweg 4-6, 14195 Berlin, Germany}

%\date{\today}

\begin{abstract}

Molecules containing a lanthanide atom have sets of electronic states arising from excitation of an inner-shell electron. These states have received little attention, but are thought to play an important role in laser cooling of such molecules and may be a useful resource for testing fundamental physics. We study a series of inner-shell excited states in YbF using resonance-enhanced multi-photon ionisation spectroscopy. We investigate the excited states of lowest energy, 8474, 9013 and 9090~cm$^{-1}$ above the ground state, all corresponding to the configuration 4f$^{13}$6s$^{2}$\,${}^{2}\!F_{7/2}$ of the Yb$^+$ ion. They are metastable, since they have no electric dipole allowed transitions to the ground state. We also characterize a state at 31050~cm$^{-1}$ that is easily excited from both the ground and metastable states, which makes it especially useful for this spectroscopic study. Finally, we study two states at 48720~cm$^{-1}$ and 48729~cm$^{-1}$, which are above the ionization limit and feature strong auto-ionizing resonances that prove useful for efficient detection of the molecules and for identifying the rotational quantum number of each line in the spectrum. We resolve the rotational structures of all these states and find that they can all be described by a very simple model based on Hund's case (c). Our study provides information necessary for laser slowing and magneto-optical trapping of YbF, which is an important species for testing fundamental physics. We also consider whether the low-lying inner-shell states may themselves be useful as probes of the electron's electric dipole moment or of varying fundamental constants, since they are long-lived states in a laser-coolable molecule featuring closely-spaced levels of opposite parity.

\end{abstract}

\maketitle

\section{\label{sec:level1}Introduction}

Atoms and ions in the lanthanide family have a rich spectrum of excited energy levels, one set arising from excitation of a 6s electron and another set due to excitation of a 4f electron. The two sets of levels are almost decoupled because transitions between them are strongly forbidden. These forbidden transitions are important because they serve as clock transitions and as excellent testing grounds for physics beyond the Standard Model. Consider the Yb$^+$ ion for example. The ground state is 4f$^{14}$6s\,${}^{2}\!S_{1/2}$ while the lowest-lying excited state is 4f$^{13}$6s$^{2}$\,${}^{2}\!F_{7/2}$. The electric-octupole transition between these states is the basis of an outstanding optical clock~\cite{Huntemann2012} and is being used to search for new physics via isotope shift measurements~\cite{Hur2022}, tests of local Lorentz invariance~\cite{Furst2020}, and searches for variations of the fundamental constants~\cite{Lange2021}. Inner-shell transitions have also been studied in neutral Er, Dy and Tm~\cite{Patscheider2021, Petersen2020, Golovizin2019}.

Similar sets of inner-shell excited states must exist in {\it molecules} that contain atoms from the lanthanide family, but very little is known about these states which have not received much attention so far. They may be of interest for fundamental physics as molecules frequently offer higher sensitivity to new physics than atoms, especially for exploring variations of fundamental constants~\cite{Kondov2019, Barontini2022}, and for measuring electric dipole moments (EDMs) of electrons and protons~\cite{Hudson2011, Andreev2018, Roussy2023, Grasdijk2021} or magnetic quadrupole moments (MQMs) of nuclei~\cite{Ho2023} that are highly sensitive probes of physics beyond the Standard Model~\cite{Safronova2018,Hutzler2020b}. Excellent examples are the YbF and YbOH molecules which are great systems for searching for new physics and have been laser cooled~\cite{Lim2018, Alauze2021, Augenbraun2020}.  A new generation of ultra-precise measurements based on such ultracold molecules are currently being prepared~\cite{Fitch2020b, Aggarwal2018, Anderegg2023}. 

In YbF, the ground state, $\text{X}\, ^{2}\Sigma^{+}$, correlates to the 4f$^{14}$6s configuration of Yb$^+$, and the series of excited states $\text{A}\,^{2}\Pi$, $\text{B}\,^{2}\Sigma^{+}$ etc. arise from excitation of the 6s electron.  However, the lowest-lying electronically excited state correlates to the 4f$^{13}$6s$^{2}$ configuration of Yb$^+$ and is the first of a series of such states that we call the ``4f hole'' states. They are important in laser cooling because the laser cooling cycle $\text{A} ^{2}\Pi_{1/2} \leftrightarrow \text{X}^{2}\Sigma^{+}$ has a leak to the low-lying 4f-hole states. The branching ratio for this leak is predicted to be about $5 \times 10^{-4}$~\cite{Zhang2022} which is too large for effective radiation pressure slowing or magneto-optical trapping of YbF. To make progress, it is necessary to address this leak, so a good understanding of the 4f hole states is needed. Recent theoretical studies~\cite{Zhang2022, Pototschnig2021} have shed some light on these states, and their approximate energies have been determined by studying the fluorescence from high-lying states into the 4f hole states using a grating spectrometer~\cite{Persinger2022}, but there has been no previous laser spectroscopy of these states.

\begin{figure}
\includegraphics[width=\columnwidth, keepaspectratio]{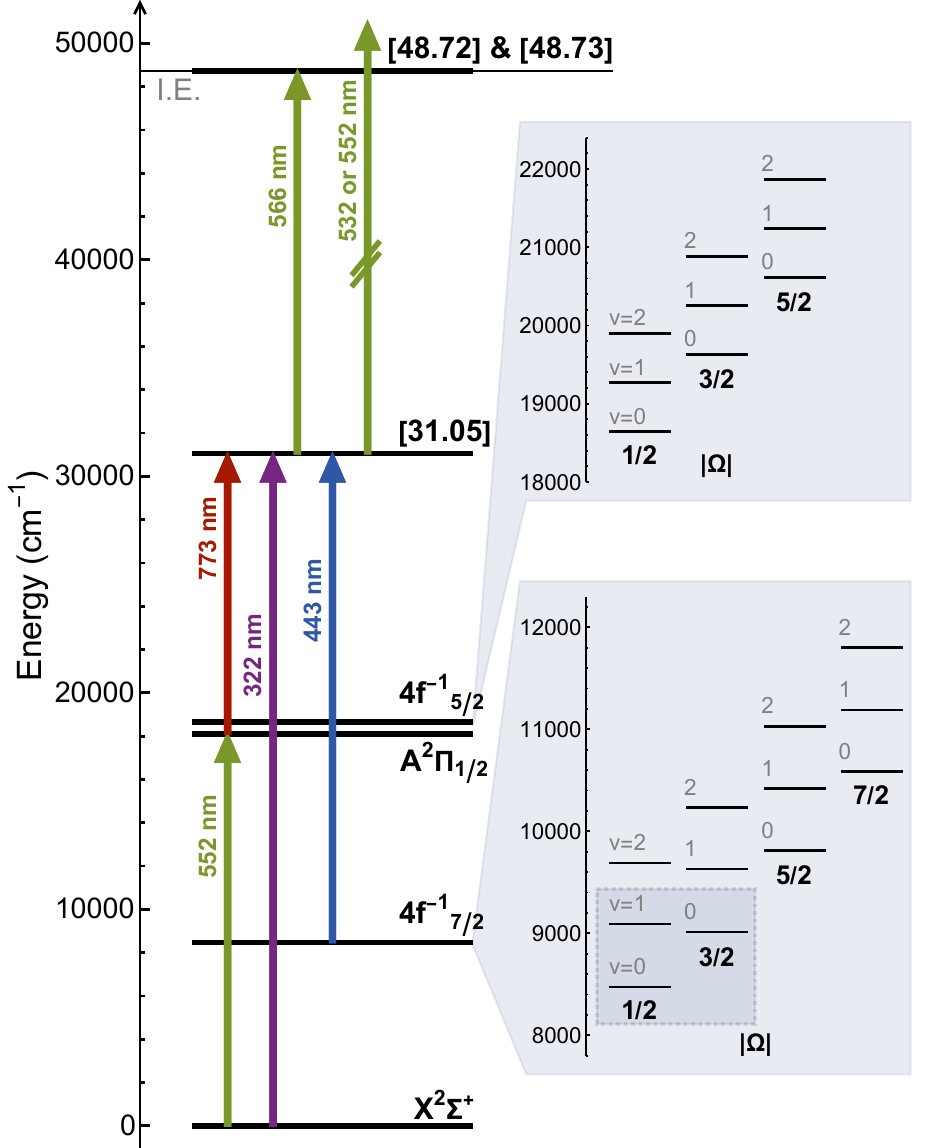}
\caption{\label{fig:OverviewEnergy} $^{174}$YbF energy levels and transitions relevant to this work. Pulsed dye lasers are used to drive the transitions shown here. A pulsed 532~nm Nd:YAG laser can in addition be used for ionisation. Light grey boxes show the structure of the $4f$ hole states, and darker grey box indicates the $4f^{-1}_{7/2}$ states studied in this work.}
\end{figure}

Here, we present rotationally-resolved spectroscopy of a series of electronic states in YbF that have 4f hole character. We begin, in Sec.~\ref{sec:experiment}, with a description of the energy levels and transitions we study and of the resonance enhanced multi-photon spectroscopic methods we use. Then, in Sec.~\ref{sec:model}, we introduce  a simple and intuitive physical model for describing these states. In this model, a 4f hole state of the Yb$^+$ ion has a Stark splitting due to the electric field from the nearby F$^-$ ion, producing a set of states characterized by the projection of the atomic angular momentum onto the electric field axis. Each of these states has a simple rovibrational structure. The atomic spin-orbit splitting is large, the Stark splitting much smaller and similar to the vibrational splitting, and the rotational splitting much smaller again. This hierarchy conforms to a Hund's case (c) description~\cite{Veseth1973}, which is rarely used in molecular spectroscopy but is simpler and more intuitive than the usual descriptions of molecular states. All the states we study fit very well to this model. 

There is no direct spectroscopic pathway to the 4f hole states from the ground state or any of the excited states arising from excitation of the 6s electron, as transitions between the two manifolds are strongly forbidden. Fortunately, there is a high-lying electronic state of mixed character, denoted [31.05], that connects strongly to both the ground state and the 4f hole states. In Sec.~\ref{sec:31.05} we characterize [31.05] using direct excitation from the ground-state and excitation from the ground-state via the $\text{A}^{2}\Pi_{1/2}$ state. We obtain a complete understanding of the rotational structure of [31.05], which is key to interpreting the spectrum obtained when exciting from the 4f hole states to [31.05]. This spectrum is complicated, so to help decipher it we devise a fingerprinting method that, for each spectral line, identifies the rotational angular momentum and parity of the upper state. The method exploits a remarkable set of very strong resonances {\it above} the ionization limit of the molecule, corresponding to auto-ionizing states that have unusually long lifetimes. We attribute the long lifetime to the states being 4f hole states that are only weakly coupled to the ground state of the YbF$^{+}$ cation, which has a filled 4f shell. These strong auto-ionizing states are themselves an interesting topic for study, and we characterize them in Sec.~\ref{sec:48.73}. 

Section \ref{sec:4fhole} presents the spectroscopy of the three 4f hole states of lowest energy. We use the information obtained in Secs.~\ref{sec:31.05} and \ref{sec:48.73} to assign quantum numbers to each line in these spectra, and thus determine their rotational level structures and their absolute energies. We show how these conform to the model introduced in Sec.~\ref{sec:model}, and note near degeneracies between levels of the same parity and between levels of opposite parity. In Sec.~\ref{sec:18.58} we look briefly at a pair of higher-lying states arising from an accidental degeneracy of a 4f hole level with a level of $\text{A}^{2}\Pi_{1/2}$, resulting in states of completely mixed character. We show that these states also conform to our model.

Finally, in Sec.~\ref{sec:Conclusions}, we discuss the implications of this work and draw our conclusions. We compare our results with calculations of the properties of the 4f hole states. We discuss a specific repumping scheme for addressing leaks to the 4f hole states, so that magneto-optical trapping of the molecule can be achieved. We also consider potential applications of the 4f hole states for tests of fundamental physics and applications of the auto-ionizing resonances for efficient detection of the molecules.

\section{Experiment}
\label{sec:experiment}

Figure \ref{fig:OverviewEnergy} presents the energy levels and transitions relevant to this work. The main optical cycling transition in YbF is between $\text{X}^2\Sigma^{+}$ and $\text{A}^2\Pi_{1/2}$. The lowest energy 4f hole states correlate to the 4f$^{13}$6s$^2$ configuration of Yb$^+$, which has  orbital angular momentum $L=3$ and spin $S=1/2$. There is a very large spin-orbit interaction that splits this configuration into a pair of states with total atomic angular momentum $J_a = 5/2$ and $7/2$, with $J_a=7/2$ having lower energy. Each of these states is further split by the electric field of the F$^-$ ion into a set labelled by the projection of the total angular momentum onto the internuclear axis, $\Omega$. We use the notation $4\text{f}^{-1}_{J_a,|\Omega|}$ to label these states, omitting $|\Omega|$ when we want to refer to the entire manifold of states. The manifold $4\text{f}^{-1}_{7/2}$ lies about half way between $\text{X}^{2}\Sigma^+$ and $\text{A}^2\Pi_{1/2}$.  The manifold $4\text{f}^{-1}_{5/2}$ lies close to $\text{A}^2\Pi_{1/2}$. Indeed, $\text{A}^2\Pi_{1/2} (v=1)$ is so close in energy to $4\text{f}^{-1}_{5/2,1/2} (v=0)$ that the pair mix almost completely~\cite{Zhang2022}. 

We also use three higher-lying states of the molecule. The first was identified by Persinger et al. \cite{Persinger2022} who called it [31.05], where the label is the term energy in thousands of cm$^{-1}$. We will continue to use this notation. This state is convenient because it has strong transitions from both the ground state (at 322~nm) and from the $\text{4f}^{-1}_{7/2}$ states (near 443~nm).  The remaining high-lying states are newly identified states at 48720~cm$^{-1}$ and 48729~cm$^{-1}$, both less than  30~cm$^{-1}$ above the ionisation energy. Continuing with the same notation, we label these as [48.72] and [48.73].

\begin{figure}
\includegraphics[width=\columnwidth, keepaspectratio]{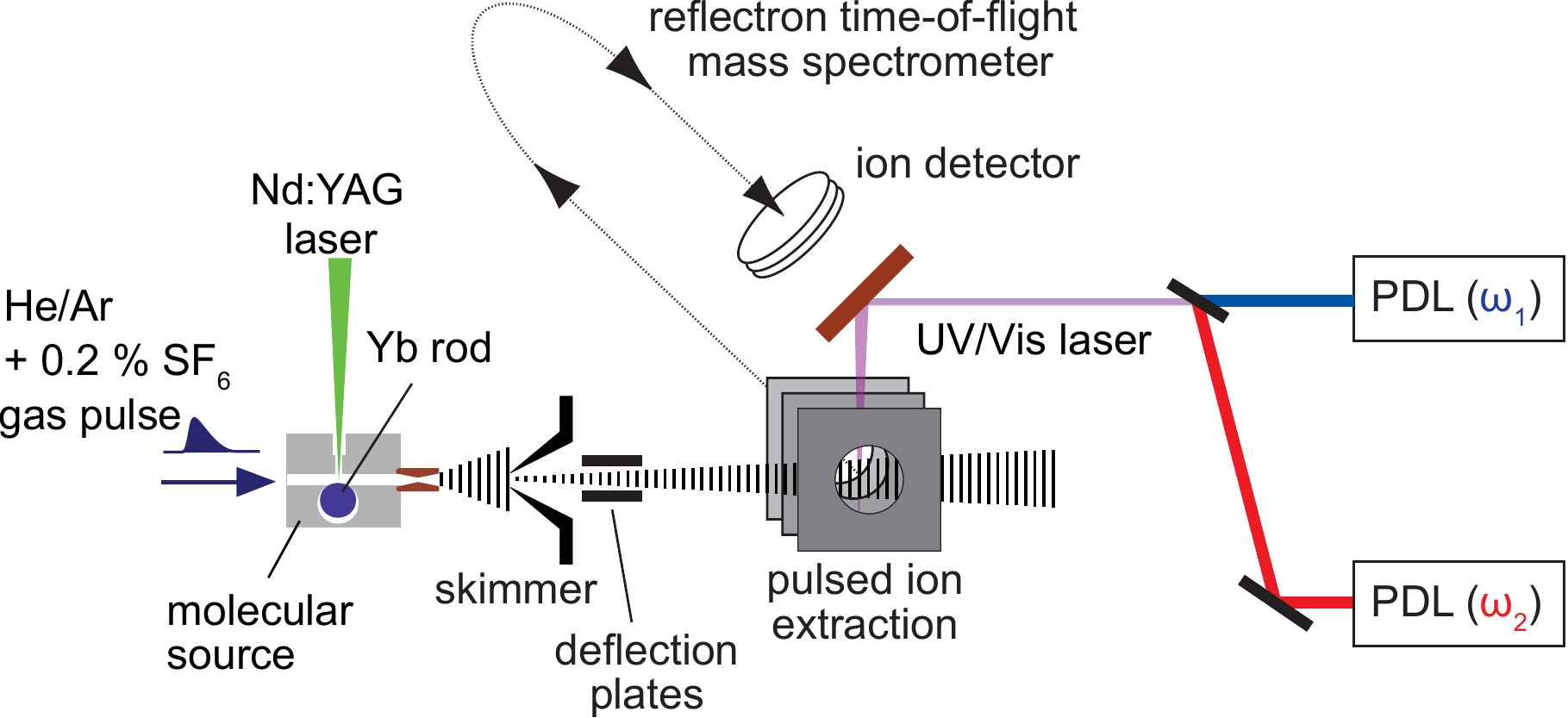}
\caption{\label{fig:Setup} Schematic of the experimental setup. YbF molecules are produced by laser ablation of a Yb rod of natural isotopic abundance in the presence of a mixture of $\text{SF}_6$ and a carrier gas. The gas mixture is introduced via a pulsed valve. The molecules interact with pulsed dye lasers (PDLs). Ion detection is carried out by pulsed ion extraction and then measured on a time-of-flight mass spectrometer in either reflectron or linear configuration. 
}
\end{figure}

Figure \ref{fig:Setup} illustrates the experiment~\cite{Fielicke2005}. We produce a supersonic beam of YbF and characterize the states of interest using resonance enhanced multi-photon ionisation (REMPI) spectroscopy. The molecules are generated by pulsed laser ablation of a solid Yb rod (natural isotopic abundance) in the presence of  $\text{SF}_6$ diluted in either a He or Ar carrier gas. The gas mixture is released into the source using a pulsed valve (R.M. Jordan Co., Inc.) operating at 10~Hz. The resulting supersonic beam first passes through a skimmer, then between a pair of plates where an applied electric field deflects away any ions generated by ablation that remain in the beam, and finally into the REMPI detection region. We find that the beam has significant population in the first $\sim$18 rotational levels of $\text{X} ^{2}\Sigma^{+}$, corresponding to a rotational temperature of $\sim$40~K. We also find that there is substantial population in the $\text{4f}^{-1}_{7/2}$ states.  

Two pulsed dye lasers are used during this study: a Sirah PrecisionScan laser with a linewidth of 0.04~cm$^{-1}$ and a Radiant Dyes NarrowScan laser with a linewidth of 0.05~cm$^{-1}$. Their frequencies are measured with a wavemeter (HighFinesse WS6-600, absolute accuracy 600 MHz). In addition, a 532 nm pulsed Nd:YAG laser can be used for ionisation. The lasers interact with the molecules in a field-free environment. In most cases the excitation laser pulse energy was such that there was little or no spectral broadening and transition widths were limited by the laser linewidths. The ionisation laser pulse energy was set low enough that no one-colour multi-photon ionisation was observed. After the laser pulses, a field is turned on to extract the ions. The cations are detected in a Wiley-McLaren type time-of-flight mass spectrometer whose resolution is sufficient to resolve the different YbF isotopologues. Although data for the various isotopologues has been obtained, we limit our discussion to the most abundant species and that most relevant for laser cooling, $^{174}$YbF.

\section{Spectroscopic model}
\label{sec:model}

\begin{figure}
\includegraphics[width=\columnwidth, keepaspectratio]{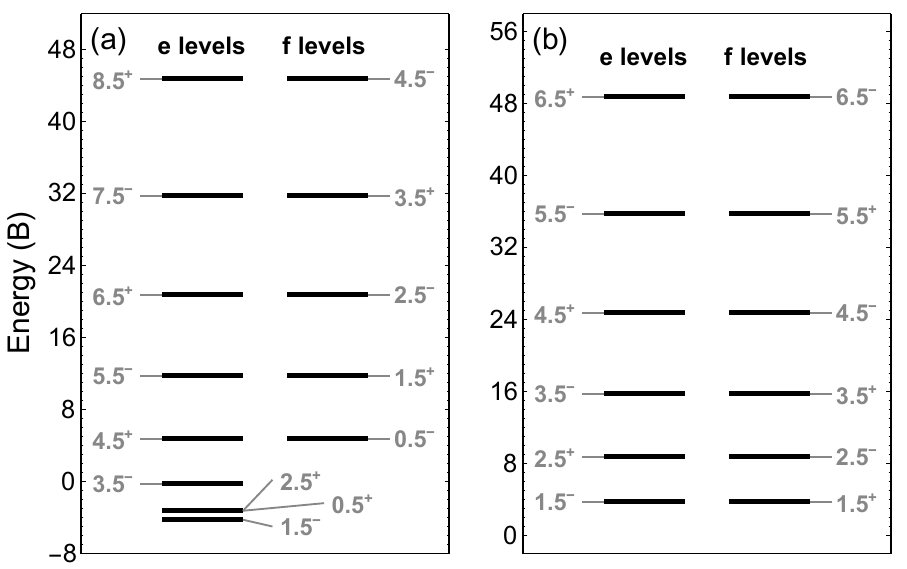}
\caption{\label{fig:case(c)example} Rotational energy levels in Hund's case (c) for the case where $J_a=7/2$ and (a) $|\Omega|=1/2$ and (b) $|\Omega|=3/2$. Levels are labelled by the total angular momentum $J$ and the parity $p$ using the notation $J^p$.\label{fig:Hundc}}
\end{figure}

\begin{table*}[tb]
\caption{\label{tab:params}Spectroscopic parameters of the states studied in this work. The parameters of [31.05] are derived from a fit to the $[31.05] \leftarrow \text{X}^2\Sigma^{+}(v=0)$ spectrum. The  parameters of [48.72] and [48.73] are derived by fitting the energy levels to Eq.~(\ref{eq:Eef2}). The parameters of the 4f$^{-1}$ states are from a fit of [31.05] $\leftarrow$ 4f$^{-1}_{7/2,|\Omega|}(v)$ spectra. The parameters of [18.58] and [18.71] are derived by fitting the data from Table 4 of Ref.~\cite{Lim2017} to Eq.~(\ref{eq:Eef2}). All energies presented in this paper are relative to the lowest rotational state of $\text{X} ^2\Sigma^+$~\cite{Sauer1996}.}
\begin{ruledtabular}
\begin{tabular}{ccccc}
Label & State composition & $T$ (cm$^{-1}$) & $B$ (cm$^{-1}$) & $\zeta$\\
\hline
$[8.47]$ & $\text{4f}^{-1}_{7/2,1/2}(v=0)$ & 8474.390(9) & 0.26737(6) & 3.8760(24) \\
$[9.01]$ & $\text{4f}^{-1}_{7/2,3/2}(v=0)$ & 9012.541(6) & 0.26597(8) & --\\
$[9.09]$ & $\text{4f}^{-1}_{7/2,1/2}(v=1)$ & 9090.675(7) & 0.2656(1) & 3.8756(24) \\
$[18.58]$ & \makecell{$\text{4f}^{-1}_{5/2,1/2}(v=0)$ \\ and $\text{A}^{2}\Pi_{1/2}(v=1)$}  & 18580.574(2) & 0.25466(6) & -1.7698(10)\\
$[18.71]$ & \makecell{$\text{4f}^{-1}_{5/2,1/2}(v=0)$ \\ and $\text{A}^{2}\Pi_{1/2}(v=1)$} & 18705.007(2) & 0.25687(6) & -1.9607(10)\\
$[31.05]$ & $|\Omega|=1/2\,(v=0)$ & 31049.541(5) &  0.24921(3) & 3.5220(12) \\
$[48.72]$ & $|\Omega|=3/2\,(v\approx4)$ & 48720.277(3) & 0.25162(3) & --\\
$[48.73]$ & $J_a=3/2, |\Omega|=1/2\,(v\approx2)$ & 48729.01(4) & 0.2568(6) & 2.016(12)\\
\end{tabular}
\end{ruledtabular}
\end{table*}

The states studied in this paper conform well to a Hund's case (c) model, as described by Veseth~\cite{Veseth1973}. We introduce the angular momenta $\vec{J}_a = \vec{L}+\vec{S}$ and $\vec{J} = \vec{J}_a + \vec{R}$ where $\vec{L}$ is the electronic orbital angular momentum, $\vec{S}$ is the electronic spin, and $\vec{R}$ is the rotational angular momentum. The projection of $\vec{J}$ (and also $\vec{J}_a$) onto the internuclear axis is $\Omega$, and the basis states are $\ket{J_a,J,\Omega}$. We may think of $\vec{J}_a$ as the angular momentum of the underlying atomic state, which in this case is the state of the Yb$^+$ ion. The state is split by the electrostatic interactions of the molecule into components labelled by $\Omega$. We think of this as the Stark splitting of the atomic state, which in this model is small compared to the spin-orbit interaction. In addition to the electronic structure, each of these states is split into smaller vibrational structure and even smaller rotational structure. The rotational structure is our main concern.

The rotational Hamiltonian is
\begin{equation}
    \begin{aligned}
    H_{\text{r}}=B\vec{R}^2=B(\vec{J}-\vec{J}_a)^2=B(\vec{J}^2+\vec{J}_a^2-2\vec{J}\cdot\vec{J}_a).
    \end{aligned}
\end{equation}
The matrix elements of $H_{\rm r}$ in the $\ket{J_a,J,\Omega}$ basis are
\begin{align}
    \bra{J_a,J,\Omega} &H_{\rm r} \ket{J_a, J, \Omega'}\!=\!B[J(J\!+\!1)\!+\!J_a(J_a\!+\!1)\!-\!2\Omega^2]\delta_{\Omega,\Omega'} \nonumber \\
    &-2B(-1)^{J_a+J-2\Omega} \sum_{q=\pm 1} g_q(J_a)g_q(J), 
    \label{eq:me_unsymm}
\end{align}
where 
\begin{equation}
    g_q(x)=\sqrt{x(x+1)(2x+1)}
    \begin{pmatrix}
    x & 1 & x\\
    -\Omega & q & \Omega'
    \end{pmatrix}.
\end{equation}
We introduce the basis set of symmetric and anti-symmetric combinations of $\Omega$, denoted by $e/f$:
\begin{equation}
    \begin{aligned}
    \ket{J_a,J,|\Omega|,^{e}_{f}} = \frac{1}{\sqrt{2}}\left(\ket{J_a,J,\Omega} \pm \ket{J_a,J,-\Omega} \right).
    \end{aligned}
\end{equation}
The diagonal elements of  $H_{\text{r}}$ in the $e/f$ basis give energies
\begin{align}
     E_{^{e}_{f}} &= BJ(J+1) + BJ_a(J_a+1) - 2B \Omega ^2 \nonumber\\ &\mp \frac{B}{4}(2J_a+1)(2J+1)\delta_{|\Omega|,1/2}.
     \label{eq:Eef}
\end{align}

The terms, $BJ_a(J_a+1)$ and $2B \Omega ^2$ are constants for each electronic state and do not contribute to the rotational structure. We incorporate these terms into the term energy, $T$. The final term in Eq.~(\ref{eq:Eef}), which depends on the e/f symmetry and is only non-zero for $|\Omega|=1/2$, comes from the off-diagonal matrix element in Eq.~(\ref{eq:me_unsymm}) which connects $\Omega=1/2$ and $\Omega=-1/2$. We see that, for $|\Omega|=1/2$, rotational level $J$ of the f manifold is degenerate with level $J+(J_a+1/2)$ of the e manifold, while for $|\Omega|>1/2$ levels of the same $J$ are degenerate in the two manifolds. There can also be degeneracy {\it within} the e manifold when $|\Omega|=1/2$: levels $J$ and $J'$ are degenerate if they satisfy $J+J'=J_a-1/2$. Figure \ref{fig:Hundc}(a) shows the energy level structure described by Eq.~(\ref{eq:Eef}) for the case where $|\Omega|=1/2$ and $J_a = 7/2$. We see the degeneracy between level $J$ of the f manifold and level $J+4$ of the e manifold. We also note the unusual behaviour of the e manifold at low $J$ -- the lowest level is $J=1.5$, while $J=0.5$ and $J=2.5$ are degenerate. Figure \ref{fig:Hundc}(b) shows the case where $|\Omega|=3/2$, where the rotational energies are simply $BJ(J+1)$ for both manifolds. All the states studied in this work have a rotational structure similar to the ideal case illustrated here, so we use this Hund's case (c) picture to describe them all.

Next, we consider two corrections that go beyond this simple model. The first is a small correction to the energy due to the off-diagonal matrix elements in Eq.~(\ref{eq:me_unsymm}) that couple states of different $|\Omega|$, differing by $\pm1$. This correction can be obtained using second order perturbation theory. For an $|\Omega|=1/2$ state, the correction is
\begin{align}
  \Delta E&=\frac{\left|\left\langle J_a, J, 1/2, ^{e}_{f}\left|H_{\text {r }}\right| J_a, J, 3/2, ^{e}_{f}\right\rangle\right|^2}{E_{1/2}-E_{3/2}}\nonumber \\&=\frac{15B^2 J(J+1) }{\Delta} - \frac{45 B^2}{4\Delta}.
\end{align}
The first term becomes a small correction to the coefficient of $J(J+1)$ in Eq.~(\ref{eq:Eef}), while the second term is a constant that we absorb into the term energy $T$. The correction is small because $\Delta = E_{1/2}-E_{3/2}$, the energy difference between the $|\Omega|=1/2$ and $|\Omega|=3/2$ states, is large compared to $B$. Although $\Delta E$ is identical for the e and f manifolds, it depends on $J$, so slightly lifts the degeneracy for $|\Omega|=1/2$ (but not for $|\Omega|>1/2$). The equivalent correction for $|\Omega|=3/2$ is
\begin{equation}
  \Delta E=\frac{15(J(J+1)-3 / 4) B^2}{E_{3 / 2}-E_{1 / 2}} + \frac{12(J(J+1)-15 / 4) B^2}{E_{3 / 2}-E_{5 / 2}} .
\end{equation}

Our second correction accounts for the fact that the molecular electric field responsible for the splitting by $|\Omega|$ can also mix configurations of different $J_a$, resulting in eigenstates of the form 
\begin{equation}
    \ket{\psi} = \sum_{J_a} c_{J_a}\ket{J_a,J,|\Omega|,^{e}_{f}},
\end{equation}
where the sum runs over the possible values of $J_a$, and $\sum_{J_a} |c_{J_a}|^2=1$. Since $H_{\rm r}$ does not couple states of different $J_a$, the only modification needed is to replace $\frac{1}{2}(2J_a+1)$ by $\zeta = \sum_{J_a} \frac{1}{2}(2J_{a}+1)|c_{J_a}|^2$ in the last term of Eq.~(\ref{eq:Eef}).

Combining all of the above into a single expression, the energy of an $|\Omega| = 1/2$ state can be written as
\begin{equation}
E_{^{e}_{f}} = T + \left( B +  \frac{15B^2}{\Delta}\right)J(J+1)  \mp \frac{B}{2}\zeta(2J+1).
\label{eq:Eef2}
\end{equation}
The last term in Eq.~(\ref{eq:Eef2}) lifts the degeneracy between the e and f manifold noted above and illustrated in Fig.~\ref{fig:Hundc}(a). It is equivalent to a lambda-type doubling term familiar from case (a) states, but here it is fully characterised by the rotational constant $B$ and the wavefunction amplitudes $|c_{J_a}|^2$. Fitting to this model will return an optimal value of $\zeta$, but the combination of $(2J_{a}+1)|c_{J_a}|^2$ factors that give a certain $\zeta$ will, in general, not be unique.

\section{[31.05]}
\label{sec:31.05}

We begin by analyzing the rotational structure of [31.05]. This is a state of mixed character which provides a convenient pathway between the ground state and the 4f hole states~\cite{Persinger2022}. Later, we use the structure of [31.05] in our analysis of the strong transition $[31.05] \leftarrow \text{4f}^{-1}_{7/2}$.

\begin{figure*}
\includegraphics[width=\textwidth, keepaspectratio]{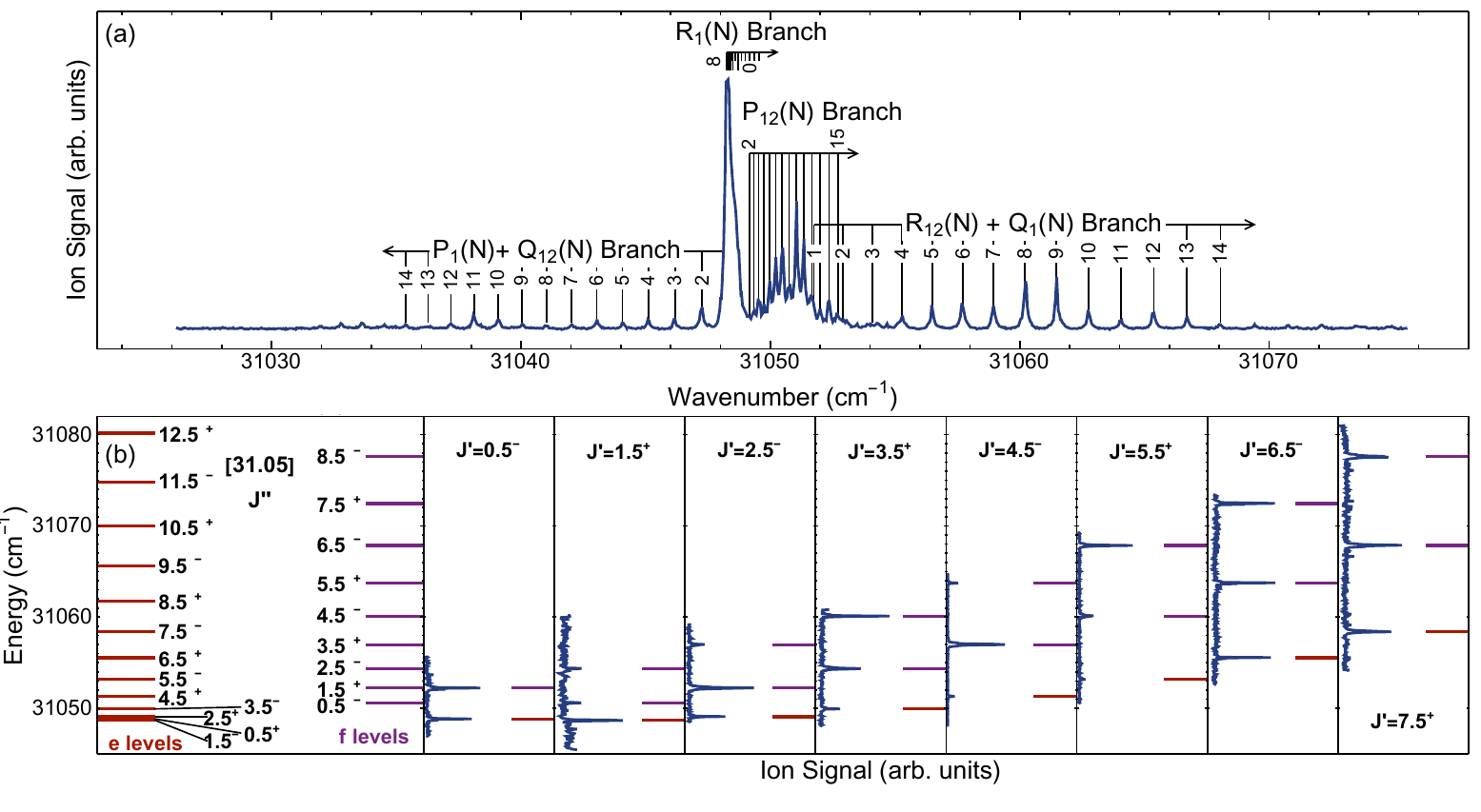}
\caption{\label{fig:31.05} Study of [31.05].  (a) $[31.05] \leftarrow \text{X}^2\Sigma^{+}(v=0)$ spectrum, with assignments of rotational branches. $N$ labels the rotational level of X. For each $N$, the total angular momentum in X can either be $J=N+1/2$ or $J=N-1/2$. P$_1$ means $\Delta J = J_{\rm upper} - J_{\rm lower} = -1$ and $J=N+1/2$. Q$_1$ means $\Delta J = 0$ and $J=N+1/2$. R$_1$ means $\Delta J = 1$ and $J=N+1/2$. P$_{12}$ means $\Delta J = -1$ and $J=N-1/2$. Q$_{12}$ means $\Delta J = 0$ and $J=N-1/2$. R$_{12}$ means $\Delta J = 1$ and $J=N-1/2$. (b) Spectra of $[31.05] \leftarrow \text{A}^2\Pi_{1/2}(v'=0,J')$, where a single value of $J'$ is used in each case and is determined by selective excitation on a known rotational transition from the X state. Each panel is labelled according to the value of $J'$, with its parity indicated as a superscript. The vertical axis is the total excitation energy relative to the lowest energy level of $\text{X} ^{2}\Sigma^+$. For each spectrum, this is obtained by adding up the energy of the initial rotational state of $\text{X} ^{2}\Sigma^+$, the known energy of the $\text{A} ^2\Pi_{1/2}\leftarrow\text{X}^2\Sigma^+$ excitation and the measured energy of the 773~nm photon, so that the vertical axis is the absolute energy. From these spectra we determine the rotational energy levels of [31.05], which are shown in the left panel and also as horizontal lines in the experimental spectra.} 
\end{figure*}

Figure \ref{fig:31.05}(a) presents the spectrum of the $[31.05] \leftarrow \text{X}^2\Sigma^{+}(v=0)$ band obtained by scanning a laser at 322 nm. The laser is a RadiantDyes NarrowScan with DCM dye producing 644 nm light which is then frequency doubled. A second pulsed dye laser at 552 nm is used to ionise from [31.05]. There is a 10~ns delay between the two laser pulses. Each point in the scan is comprised of the ion intensity extracted from the average of 30 consecutive mass-spectra traces. 

Assignment of the spectral lines proved challenging because of the congestion of lines in the central region of the spectrum together with the irregular intensities observed. Instead, assignment of the rotational lines was made by excitation via a well characterised intermediate state. The first step is to excite molecules to a known rotational level $J'$ of $\text{A}^2\Pi_{1/2}(v=0)$.  
We do this using a pulsed dye laser tuned to a selected rotational line in either the $\text{P}_{12}$ or $\text{R}_{1}$ branches of the $\text{A}^2\Pi_{1/2}(v=0) \leftarrow \text{X}^2\Sigma^{+}(v=0)$ transition at 552~nm (branch notation is the same as described in the caption of Fig.~\ref{fig:31.05}). The rotational structure of these branches is well resolved and well characterized, so we can be sure to populate only one rotational state, $J'$. Then, we use a second pulsed dye laser at 773~nm to drive $[31.05] \leftarrow \text{A}^2\Pi_{1/2}(v=0,J')$ transitions. There are, at most, 3 possible transitions, due to the angular momentum selection rule, so the spectrum is hugely simplified. We use a 532~nm pulsed Nd:YAG laser to ionise from [31.05], and then detect these ions. There is a 10~ns time delay between each of the laser pulses. 

Figure \ref{fig:31.05}(b) shows scans of the $[31.05] \leftarrow \text{A}^2\Pi_{1/2}(v=0,J')$ transition for various values of $J'$. In each case, we add together the measured energy of the 773~nm photon with the known energy of the $\text{A} ^2\Pi_{1/2}\leftarrow\text{X}^2\Sigma^+$ excitation and the known energy of the initial rotational level of $\text{X}^2\Sigma^+$, so that the vertical axis is the absolute energy. As expected, the method produces very simple and unambiguous spectra.  For example, excitation of the P$_{12}$(2) line prepares the negative parity component with $J'=0.5$ of $\text{A}^2\Pi_{1/2}(v=0)$. From here the only levels of [31.05] that can be excited are positive parity components with $J''=0.5$ and 1.5. We see only these two lines in the spectrum. The $J''=1.5$ line (but not $J''=0.5$) is observed again when we excite from the negative parity component of $J'=2.5$, providing an unambiguous assignment. Continuing in this way, all lines can be assigned. Our method effectively measures the rotational energy levels directly, rather than the usual method of extracting the level structure from a fit to a spectrum.
 
The rotational energy levels determined this way are shown on the left of Fig.~\ref{fig:31.05}(b). We immediately see the similarity to the structure illustrated in Fig.~\ref{fig:Hundc}(a), clearly identifying that [31.05] is a state with $|\Omega|=1/2$. We see that level $J$ of the f manifold lies between levels $J+3$ and $J+4$ of the e manifold. A preliminary fit of these energy levels is used in order to assign the observed transitions in Fig.~\ref{fig:31.05}(a). We then fit a Lorentzian model to individual rotational lines in the spectrum where these do not overlap significantly. Finally, a non-weighted, non-linear fit of 43 lines from this spectrum is used to determine the spectroscopic constants. The result of this fit is shown in Table \ref{tab:params}. The standard deviation of the residuals is 0.02~cm$^{-1}$. We assign this transition as a 0-0 vibrational band since we do not observe isotope shifts between the rotational spectra of the five dominant isotopes. The value of $\zeta=3.522$ corresponds to the case where the state is a 52:48 mixture of two electronic states with $J_a=7/2$ and $J_a=5/2$.  

\begin{figure}
\includegraphics[width=\columnwidth, keepaspectratio]{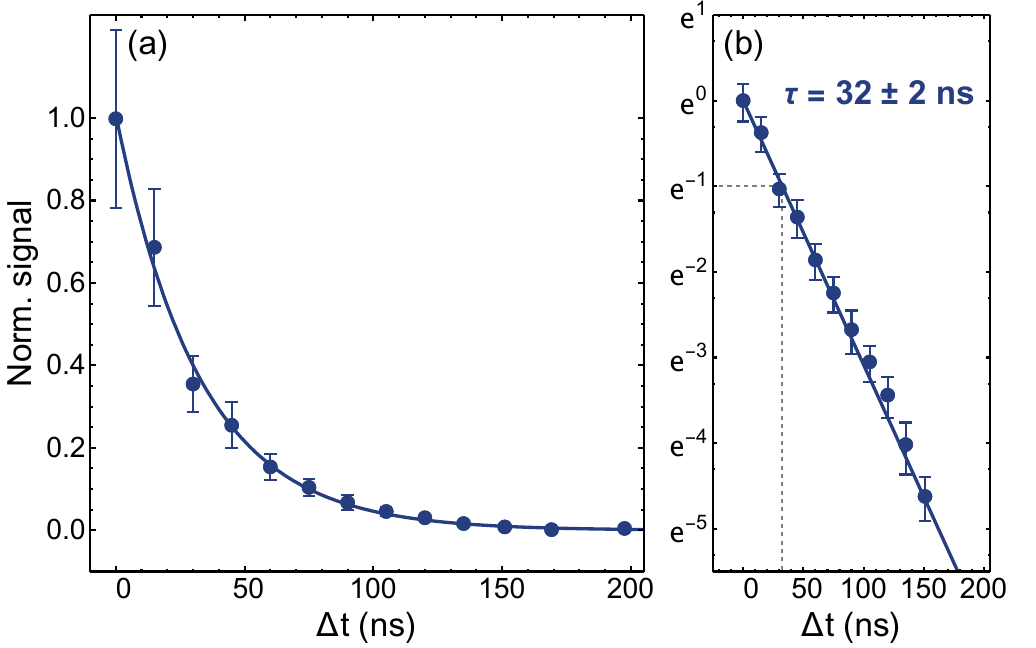}
\caption{\label{fig:31.05lifetime} Lifetime measurement of [31.05]. (a) Normalised ion signal as a function of the
time delay, $\Delta t$, between the excitation and ionisation laser pulses. The line is a fit to a single exponential decay. (b) The same data presented on a natural log scale. The error bars indicate the standard error on each measurement point.}
\end{figure}

The lifetime of [31.05] is determined by time-resolved photo ionisation using the same REMPI scheme as used to obtain the data in Fig.~\ref{fig:31.05}(b). The ion signal is recorded as a function of the time delay between the 773~nm excitation and 532~nm ionisation laser pulses. We made this measurement for six different rotational states of [31.05] and saw no significant difference between them. Figure \ref{fig:31.05lifetime} shows the mean of these six measurements, along with a fit to a single exponential model. The lifetime is 32(2)~ns. 

\section{[48.72] and [48.73]}
\label{sec:48.73}

We use [31.05] as the upper state in our spectroscopy of the 4f hole states, and the information obtained in the last section is crucial for understanding these spectra. Even with this information, the spectra were difficult to interpret. The relative intensities of rotational lines can be a great help in assigning quantum numbers in a spectrum, particularly when rotational lines overlap. In the present study, however, the observed intensities do not follow a thermal distribution or the expected H\"onl-London factors as is already evident from the spectrum shown in Fig.~\ref{fig:31.05}.  Instead, the observed intensity pattern is skewed because we reach auto-ionizing resonances above the ionisation energy of the molecule. These resonances prove to be both interesting and useful for understanding the 4f hole spectra. In this section we study these auto-ionizing resonances, and in the next section we make use of them to help interpret the 4f hole spectra.

\begin{figure}
\includegraphics[width=\columnwidth, keepaspectratio]{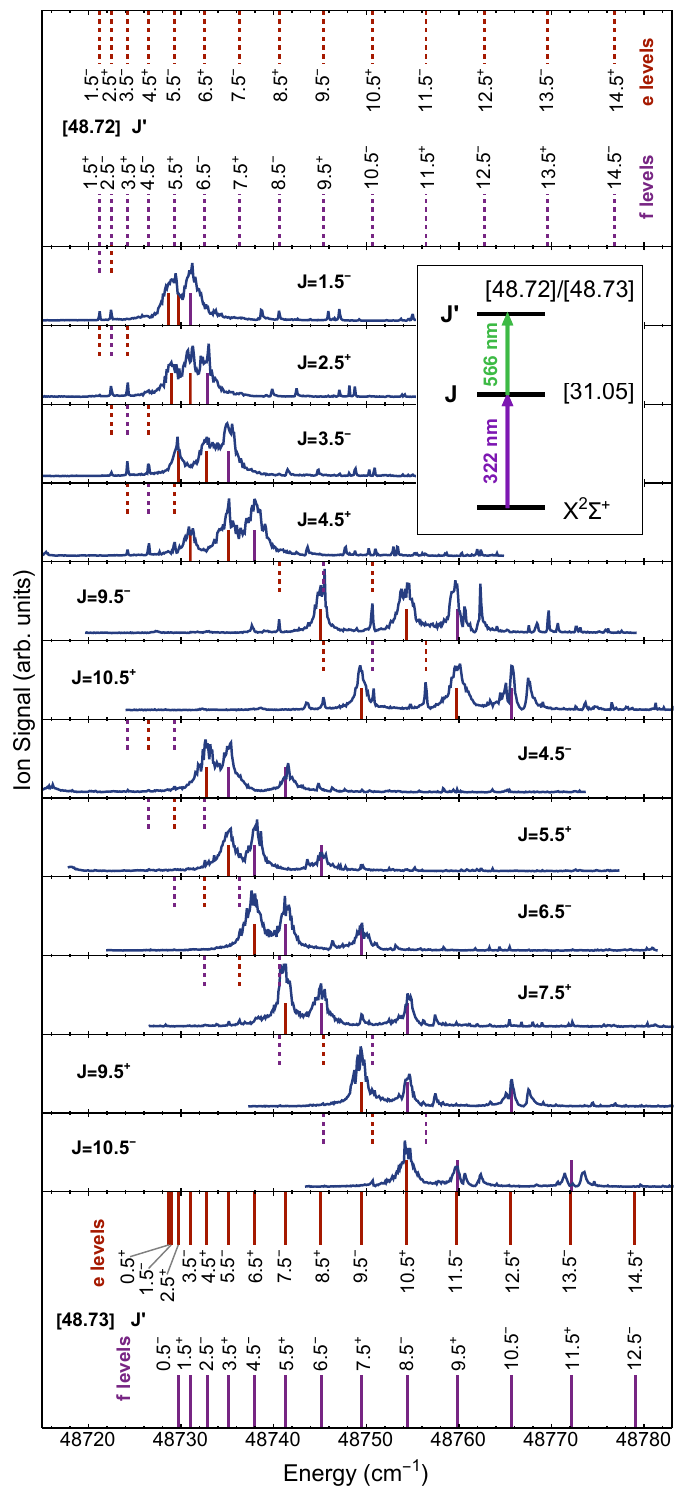}
\caption{\label{fig:48.73} Spectroscopic study of [48.72] and [48.73]. The inset shows the selective ionization scheme used. The first step prepares population in a single rotational level, $J$, of [31.05]. The rotational structure of [48.72] and [48.73] is then measured as the 566~nm laser is scanned. The set of middle panels show the spectra obtained for various $J$ and for both parity components. The horizontal axis gives the total energy. The fitted energy levels are shown as red/purple vertical lines (dashed for [48.72], solid for [48.73]). The upper and lower panels show the complete set of energy levels determined by this spectroscopy.}
\end{figure}

The ionization energy of YbF is at $48703\pm5~\text{cm}^{-1}$~\cite{Persinger2022}. Figure \ref{fig:48.73} shows scans above this ionisation energy. The excitation sequence is illustrated in the inset. First, a single rotational state ($J$) of [31.05] is populated by driving a single line of either the $\text{P}_{1}\left( N \right)$ or $\text{R}_{12}\left( N \right)$ branches of the $\text{[31.05]} \leftarrow \text{X}^2\Sigma^{+}(v=0)$ transition (see Fig.~\ref{fig:31.05}(a)). Then, a second laser at around 566~nm is scanned, resulting in ionization of the molecule. The horizontal axis in Fig.~\ref{fig:48.73} is the total energy relative to the lowest level of $\text{X}^2\Sigma^{+}$. We determine this for each spectrum by adding together the known energy of the initial level of X, the fitted energy of $[31.05]\leftarrow\text{X}^2\Sigma^{+}$ and the energy of the 566~nm photon. The spectra in Fig.~\ref{fig:48.73} contain a sequence of narrow resonances with Lorentzian widths limited by the laser linewidth, corresponding to states with remarkably long lifetimes of over 100~ps, and a series of broader resonances with Lorentzian widths (FWHM) of 1.4~cm$^{-1}$, corresponding to a lifetime of 3.7~ps. We first focus on these broad resonances and find that they correspond to the rotational structure of a state that we call [48.73] because its term energy is close to 48730~cm$^{-1}$. The uppermost spectrum in the figure corresponds to excitation from the negative parity, $J=1.5$ level in [31.05]. This was excited by driving the $\text{P}_{1}(2)$ line of the $\text{[31.05]} \leftarrow \text{X}^2\Sigma^{+}(v=0)$ transition. 
The three allowed transitions to the positive parity $J^\prime = 0.5, 1.5, 2.5$ levels are marked by vertical lines. The positive parity $J^\prime = 2.5$ level is measured again in the spectrum where the selected lower level is the negative parity $J = 3.5$ level, confirming its assignment. The energy level structure of [48.73] is discerned by repeating this process for all the values of $J$ shown in the figure. The resulting level structure is shown at the bottom of Fig.~\ref{fig:48.73}. Within our uncertainties, we find that level $J'$ of the f manifold is degenerate with level $J'+2$ of the e manifold, which tells us that [48.73] has $|\Omega|=1/2$ and $J_a = 3/2$, and that it conforms exactly to the simple Hund's case (c) model described in Sec.~\ref{sec:model}. Fitting the energy levels to Eq.~(\ref{eq:Eef2})\footnote[1]{We set the term proportional to $B^2/\Delta$ to zero in this fit} gives the parameters summarised in Table \ref{tab:params}. The standard deviation of the residuals from this fit is 0.1 cm$^{-1}$, which is commensurate with the increased uncertainty due to the width of the transitions. There is no obvious structure in the residuals. By comparing our measured spectra for the various isotopologues of YbF, we find that the rotational lines of $[48.73]\leftarrow[31.05]$ show a significant isotope-dependent shift. This isotope shift is consistent with [48.73] having about 1000~cm$^{-1}$ more vibrational energy than [31.05]. Since [31.05] has $v=0$ and vibrational constants in YbF are typically around 500~cm$^{-1}$, we suggest that [48.73] has $v\approx 2$.  

The structure of the narrow transitions can be discerned in much the same way. In keeping with the nomenclature above, we label the series of narrow peaks near 48720~cm$^{-1}$ as [48.72]. Two key observation can be made in the structure of this state. First, we see that there are no excitations $J^\prime=0.5 \leftarrow J = 1.5$,  i.e. $J^\prime = 1.5$ is the first rotational state. Second, we find that  levels with the same $J^\prime$ in the e and f manifolds are degenerate. In our Hund's case (c) model, this conforms well to a state of $|\Omega|=3/2$. The term energy, $T$, and rotational constant, $B$, obtained by a weighted fit to Eq. (\ref{eq:Eef2}) are presented in Table \ref{tab:params}. For an $|\Omega|>1/2$ state there is no $\zeta$ parameter, so only these two constants are relevant\footnotemark[1]. The residuals from this fit all fall within their uncertainties. Transitions $[48.72]\leftarrow[31.05]$ show a large isotope shift consistent with [48.72] having almost 2000~cm$^{-1}$ more vibrational energy than [31.05]. Such an isotope shift would be consistent with $v\approx 4$. We note that excitations to [48.72] via $[31.05] \leftarrow \text{X}^2\Sigma^+ (v=0)$ R$_{12}$ transitions are not as visible in Fig.~\ref{fig:48.73} as those obtained via excitation on P$_1$ transitions. 

Each ionisation spectrum in Fig.~\ref{fig:48.73} is a unique `fingerprint' of the rotational level prepared in [31.05]. We now use these fingerprints to help us characterize the $\text{4f}_{7/2}^{-1}$ states. 

\section{$\text{4f}_{7/2}^{-1}$ states}
\label{sec:4fhole}

In this section we present rotationally resolved spectra of the $[31.05] \leftarrow \text{4f}_{7/2}^{-1}$ transitions measured using two colour (1+1$^\prime$) REMPI. The supersonic source produces sufficient population in the $\text{4f}_{7/2}^{-1}$ states to do this spectroscopy directly.

\begin{figure*}
\includegraphics[width=\textwidth, keepaspectratio]{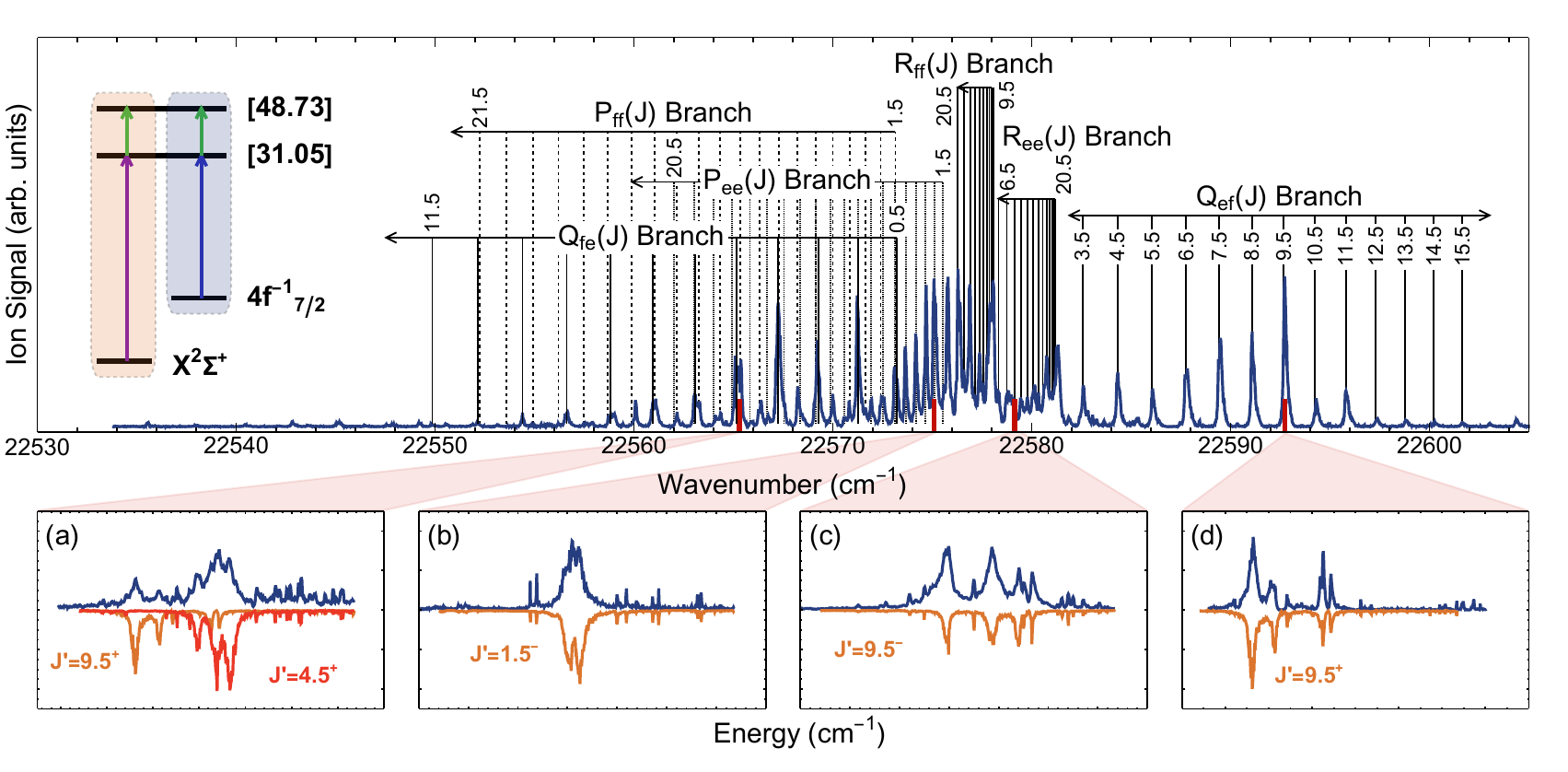}
\caption{\label{fig:4fv0}The rotational spectrum of the $[31.05] \leftarrow \text{4f}_{7/2,1/2}^{-1}(v=0)$ transition. The branch labelling is described in the text. Example ionisation spectra are shown in (a) - (d). Characterisation of the rotational branches was made by recording ionisation spectra (blue) at selected excitation frequencies (red lines in top panel) and then comparing them to a reference spectrum (orange) obtained by starting from the X$^2\Sigma^+$ state. The REMPI schemes to obtain blue and orange curves are highlighted in the energy level diagram on the left.}
\end{figure*}

Figure \ref{fig:4fv0} shows the observed spectrum of $[31.05] \leftarrow \text{4f}_{7/2,1/2}^{-1}(v=0)$. The excitation scheme used to obtain the spectrum is shaded in blue. A first excitation laser driving the $[31.05] \leftarrow \text{4f}_{7/2}^{-1}$ transition is scanned. A second `ionisation' laser drives the $[48.73] \leftarrow [31.05]$ transition, producing ions which are then extracted and detected in the time-of-flight mass spectrometer. There is a 10~ns delay between the two laser pulses. The ion signal is much stronger when detected resonantly via [48.73], than when ionizing with 532~nm light. Since the auto-ionising states are broad, a single ionisation laser frequency is sufficient to observe many rotational lines in the $[31.05] \leftarrow \text{4f}_{7/2}^{-1}$ spectrum. Note that the intensity of any given rotational line in the spectrum will depend on the strengths of the two excitation steps and also on the detuning of the ionization laser from the resonance relevant to that rotational line.  The intensity pattern can be very different if a different frequency is chosen for the ionisation laser, which makes it challenging to extract useful information from the intensities. 

\begin{figure*}
\includegraphics[width=\textwidth, keepaspectratio]{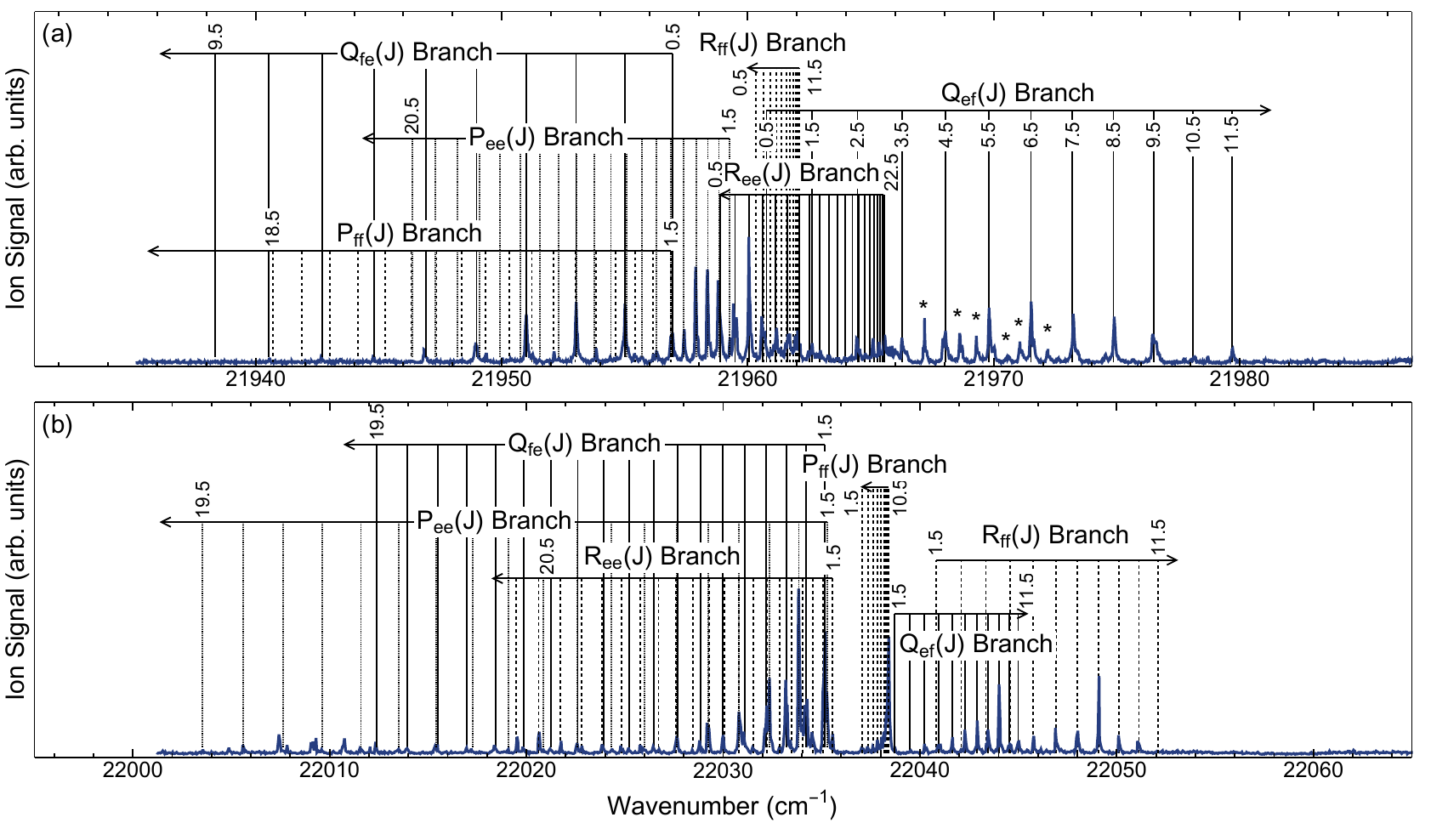}
\caption{\label{fig:4fv1andOmegaSpectra} (a) Spectrum of the $[31.05] \leftarrow \text{4f}_{7/2,1/2}^{-1}(v=1)$ transition. Note that there are a few unassigned lines near 21968~cm$^{-1}$ (marked with an *) that do not belong to this transition. (b) Spectrum of the $[31.05] \leftarrow \text{4f}_{7/2,3/2}^{-1}(v=0)$ transition. The notation is the same as in Fig.~\ref{fig:4fv0}.}
\end{figure*}

\begin{figure*}
\includegraphics[width=0.85\textwidth, keepaspectratio]{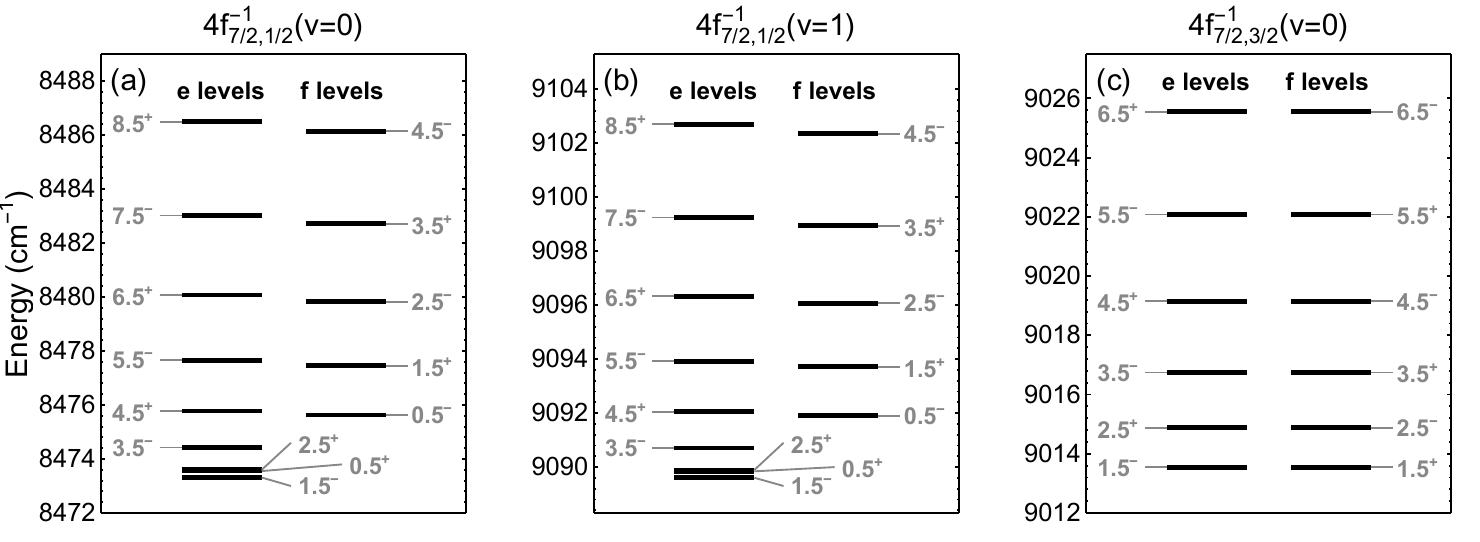}
\caption{\label{fig:4fRotationalLevels} Summary of the energy level structures of the low-lying 4f hole states studied in this work: (a) 4f$^{-1}_{7/2,1/2}(v=0)$, (b) 4f$^{-1}_{7/2,1/2}(v=1)$, (c) 4f$^{-1}_{7/2,3/2}(v=0)$. }
\end{figure*}

Assignment of the rotational lines was achieved with the help of the reference ionization spectra (or `fingerprints') noted above. We choose a particular line in the spectrum and fix the frequency of the excitation laser to the frequency of that line. This excites some particular rotational level $J'$ of [31.05]. To determine $J'$, we now scan the ionization laser across the [48.73] state. The resulting ionization spectrum is unique for each $J'$ and can be compared to the `reference' ionisation spectra in Fig.~\ref{fig:48.73}.  Figure \ref{fig:4fv0}(a-d) shows four examples of this procedure, corresponding to the four positions marked by red lines in the upper panel. The blue lines are the ionisation spectra obtained after excitation from $\text{4f}_{7/2,1/2}^{-1}(v=0)$, while the orange lines are the reference spectra from Fig.~\ref{fig:48.73}. As an example, consider Fig.~\ref{fig:4fv0}(d) where the excitation frequency is 22592.73~cm$^{-1}$. The ionisation spectrum matches that of the positive parity $J^\prime=9.5$ level in [31.05]. This identifies the upper level for this line in the spectrum, but not yet the lower level which could have $J=8.5, 9.5$ or 10.5. The same level of [31.05] is reached again when we excite at 22565.30~cm$^{-1}$, as shown in Fig.~\ref{fig:4fv0}(a). Here, the ionization spectrum is more complicated, corresponding to the sum of two fingerprints, namely the positive parity components of $J'=9.5$ (orange) and $J'=4.5$ (red), meaning that two distinct lines are excited at this frequency. Following this procedure, each rotational level $J$ of $\text{4f}_{7/2,1/2}^{-1}(v=0)$ can in principle be observed three times, corresponding to the P, Q and R lines going to $J^\prime=J-1$, $J$ and $J+1$ of [31.05]. For each rotational line observed in the spectrum the absolute energy of the lower level can be calculated by subtracting the excitation frequency from the known energy of the [31.05] rotational level we are exciting to. A unique assignment of the rotational quantum number $J$ can be made once a P and a Q line coming from the same $J$ (identified by having the same lower energy) have been measured. The energy level structure we determine this way is summarized in Fig.~\ref{fig:4fRotationalLevels}(a). The similarity to the level structure for the ideal $J_a=7/2,|\Omega|=1/2$ state is evident by comparison to Fig.~\ref{fig:Hundc}(a).

The rotational bands in Fig.~\ref{fig:4fv0} are labelled using the notation $\Delta J_{p^{\prime}p^{\prime\prime}}$ where $p^\prime$ and $p^{\prime\prime}$ are either $e$ or $f$ and denote the symmetry of the $\text{4f}_{7/2}^{-1}$ and [31.05] rotational states, respectively.  
We observe two bandheads in the R$_{ff}$ and $\text{R}_{ee}$ branches occurring at $J=9.5$ and $J=20.5$ respectively. We fit Lorentzians to the rotational lines in the spectrum. The low frequency side of the spectrum is congested with the $\text{P}_{ff}$,  $\text{P}_{ee}$ and $\text{Q}_{fe}$ lines all overlapping. To better resolve this region of the spectrum, we set the ionisation frequency to selectively ionise only some rotational levels. For example, in Fig.~\ref{fig:4fv0}(a) we see that there exist ionisation frequencies where only the $\text{P}_{ff}(10.5)$ line and $\text{Q}_{}(4.5)$ lines are respectively detected. This procedure can be repeated for many of the overlapping peaks. In total, we fit 61 lines to the model described by Eq.~(\ref{eq:Eef2}). In this fit, we fix $\Delta$ to our measured difference in term energies for $|\Omega|=1/2$ and 3/2, which is $\Delta=-538.151$~cm$^{-1}$ (see Table \ref{tab:params} and next paragraph). Table \ref{tab:params} gives the best fit parameters. The rotational constant is in agreement with the value predicted from the bond length calculated in \cite{Zhang2022}. The value of $\zeta$ does not uniquely determine the coefficients $c_{J_a}$ but is consistent with a dominantly $J_a = 7/2$ state; the upper bound on $|c_{7/2}|$ is 0.94.

Figure \ref{fig:4fv1andOmegaSpectra} shows our measured spectra  for the $[31.05] \leftarrow \text{4f}_{7/2,1/2}^{-1}(v=1)$ 
and $[31.05] \leftarrow \text{4f}_{7/2,3/2}^{-1}(v=0)$ transitions. We use the same methodology as described above to assign quantum numbers to the lines in these spectra, and the same analysis methods to determine the spectroscopic constants, using 44 and 47 lines respectively.  The energy difference, $E_{3/2}-E_{5/2}$, is set to -790~cm$^{-1}$ based on the work by Persinger et al.~\cite{Persinger2022}. For the state with $|\Omega|=3/2$ the $\zeta$ term is not relevant and we omit it. 

The spectroscopic constants we determine are reported in Table \ref{tab:params}, and the energy level structures are summarized in Fig.~\ref{fig:4fRotationalLevels}. 

\section{[18.58] and [18.71]}
\label{sec:18.58}

The $v=0$ level of $\text{4f}_{5/2,1/2}^{-1}$ lies very close in energy to the $v=1$ level of $ \text{A}^2\Pi_{1/2}$, and they are strongly mixed. In our notation, the resulting eigenstates are labelled [18.58] and [18.71]\footnote[2]{Elsewhere, e.g \cite{Lim2017,Zhang2022}, these states are instead labelled by their frequencies in THz: [557] and [561].}. These two states have strong transitions to X$^2\Sigma^+$, and their rotational energy level structure has been measured previously by cw laser spectroscopy of the $[18.58] \leftarrow \text{X}^{2}\Sigma^+$ and $[18.71] \leftarrow \text{X}^{2}\Sigma^+$ transitions~\cite{Lim2017}. The states were characterized by modelling them as $^2\Pi_{1/2}$ states in Hund's case (a). By fitting the energy levels given in Table 4 of \cite{Lim2017}\footnote[3]{There is an error in Table 4 of \cite{Lim2017}: the e and f labels are the wrong way around.} to Eq.~(\ref{eq:Eef2})\footnotemark[1], we find that both states fit well to our model of an $|\Omega|=1/2$ state in Hund's case (c).  The residuals show some structure but are all below 0.01~cm$^{-1}$. The parameters of these fits are given in Table \ref{tab:params}. For both states, the value of $B$ is consistent with the value determined from the $^2\Pi_{1/2}$ model. The values of $\zeta$ are negative, consistent with the negative value of the $\Lambda$-doubling constant ($p+2q$) when modelled as a case (a) state. A discussion of the sign of this parameter is given by Hougen~\cite{Hougen2011}.

A previous deperturbation analysis~\cite{Zhang2022} suggests that [18.58] has 52\% $ \text{A}^2\Pi_{1/2}(v=1)$ character and 46\% $\text{4f}_{5/2,1/2}^{-1}(v=0)$ character, while for [18.71] the fractions are 47\% and 51\%. The remaining composition comes from other vibrational levels of the same electronic states and they are all at least an order of magnitude smaller. Here, we study these compositions by measuring the lifetimes of the two states. The transition dipole moment for the A-X transition is about 100 times larger than for the 4f$^{-1}_{5/2,1/2}$-X transition. In this limit, where the mixing is strong and one transition dipole is much larger than the other, the lifetime of a mixed state whose A state amplitude is $c_A$ is simply $\tau_{\rm mixed} \simeq \tau_{A}/|c_A|^2$, where $\tau_{A}$ is the lifetime of the A state.

\begin{figure}
    \includegraphics[width=\columnwidth, keepaspectratio]{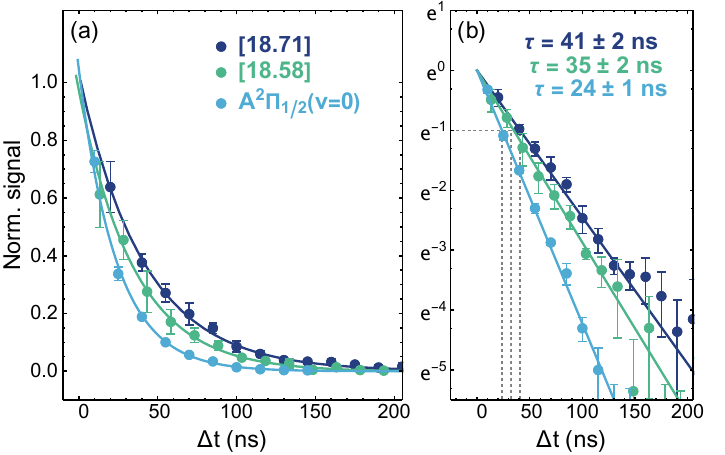}
    \caption{\label{fig:557lifetime} Lifetime measurements of [18.58], [18.71] and $\text{A}^2\Pi_{1/2}(v=0)$ states of $^{174}$YbF. The data are presented on a natural log scale in (b). The error bars indicate the standard error on each measurement point.}
\end{figure}

We have measured the lifetimes of $\text{A}^2\Pi_{1/2}(v=0)$, [18.58] and [18.71] using the delayed ionisation method described in Sec.~\ref{sec:31.05}. In each case, we measured the lifetime on three different rotational lines and averaged the data. Figure \ref{fig:557lifetime} shows our results. 
Fitting single exponential decays to these data we obtain the following lifetimes: $\tau_A = 24(1)~\text{ns}$, $\tau_{[18.58]} = 35(2)~\text{ns}$ and $\tau_{[18.71]} = 41(2)~\text{ns}$. Our value of $\tau_{A}$ is consistent with the value of $28(2)~\text{ns}$ measured previously~\cite{Zhuang2011}. These results imply that for [18.58] $|c_A|^2 = 69(5)$\%, while for [18.71] $|c_A|^2 = 58(4)$\%. These percentages are both a little larger than determined through the deperturbation analysis, but confirm that the two mixed states are indeed close to 50:50 mixtures.

\section{Conclusions and discussion}
\label{sec:Conclusions}

Our rotationally resolved spectroscopic study of the low-lying 4f hole states of YbF has determined the character of these states. They derive from the 4f$^{13}$6s$^{2}$ configuration of Yb$^{+}$, which is split by an enormous spin-orbit interaction into two terms, the one with $J_a=7/2$ lying approximately 10000~cm$^{-1}$ below the one with $J_a=5/2$. Each of these is split by the electric field of the F$^-$ ion into a set of states labelled by $|\Omega|$, but this splitting is much smaller, 500-1000~cm$^{-1}$. The rotational splitting is much smaller again, around 1~cm$^{-1}$, resulting in a convenient structural hierarchy conforming to Hund's case (c). The rotational structures of the 4f hole states fit very well to this model, requiring only 3 parameters -- a term energy, a rotational constant, and an effective value of $J_a+1/2$. We have determined these constants to high precision. In previous work, the energies, bond lengths and vibrational constants of the 4f hole states were calculated~\cite{Zhang2022}. Our measured rotational constants imply a bond length of 1.9187(2)\AA, consistent with the calculations. Similarly, the energy difference between $\text{4f}_{7/2,1/2}^{-1} (v=0)$ and $\text{4f}_{7/2,1/2}^{-1} (v=1)$, measured to be 616.285(11)~cm$^{-1}$, is close to the calculated value of 614.3~cm$^{-1}$.

We consider the splitting between the $|\Omega|$ components of the 4f hole states to be the Stark splitting of the Yb$^+$ 4f$^{13}$6s$^2\,{}^2{F}_{7/2}$ state. To support this picture, we estimate the polarizability of this configuration using second-order perturbation theory, summing over all relevant states of the ion. A table of the required transition frequencies and Einstein A-coefficients is given in \cite{Biemont1998}. We find the difference in polarizability between the $|\Omega|=1/2$ and $|\Omega|=3/2$ components to be $\alpha_{1/2}-\alpha_{3/2} = 1.19\times 10^{-41}$~J\,m$^2$\,V$^{-2}$. Our measured splitting is $\Delta = E_{1/2} - E_{3/2} = -538.15$~cm$^{-1}$, implying an electric field of $4.25 \times 10^{10}$~V\,m$^{-1}$ at the Yb$^+$ ion. This is the electric field produced by a charge placed 1.84$~\text{\AA}$ away. Given the simplicity of this picture, and the doubtful validity of using perturbation theory for such an enormous electric field, the result is remarkably close to the actual bond length of 1.92~\AA.

\begin{table}[tb]
\caption{\label{tab:repumps} Frequencies (in cm$^{-1}$) required to repump population from the relevant 4f hole states through either the [18.58] or [18.71] states. The uncertainty is mainly limited by the absolute accuracy of the wavemeter and is less than 0.03~cm$^{-1}$}
\begin{ruledtabular}
\begin{tabular}{ccc}
 & $[18.58](J=1/2,e)$ & $[18.71](J=1/2,e)$ \\
\hline
$[8.47](J=1/2,f)$ & 10105.609 & 10230.091 \\
$[8.47](J=3/2,e)$ & 10107.956 & 10232.438\\
$[9.09](J=1/2,f)$ & 9489.315 & 9613.797\\
$[9.09](J=3/2,e)$ & 9491.612 & 9616.094\\
\end{tabular}
\end{ruledtabular}
\end{table}

YbF, and similar molecules, are important for testing fundamental physics through measurements of the electron's electric dipole moment (EDM). The precision of these measurements can be improved by cooling the molecules to $\mu$K temperatures and trapping them~\cite{Fitch2020b}. Previous work~\cite{Alauze2021,Zhang2022} suggests that laser slowing and magneto-optical trapping using the $\text{A}^{2}\Pi_{1/2} \leftrightarrow \text{X}^{2}\Sigma^+$ transition will be hindered by small leaks from $\text{A}^{2}\Pi_{1/2}$ into $\text{4f}_{7/2,1/2}^{-1}(v=0)$ and $\text{4f}_{7/2,1/2}^{-1}(v=1)$. These small leaks open up because of a small mixing between the $\text{4f}_{7/2,1/2}^{-1}$ and $\text{X}^2\Sigma^+$ states. Since the mixing preserves $\Omega$, leaks to the 4f hole states with $|\Omega|>1/2$ are thought to be negligible. Leaks to 4f hole states with $v>1$ are also sufficiently small due to small Franck-Condon factors. Given the upper state used for laser cooling, and the parity and angular momentum selection rules, leaks should only populate the negative parity components with $J=0.5$ and $J=1.5$. This population can be returned to the laser cooling cycle using additional repump lasers. Repumping via $\text{A}^{2}\Pi_{1/2}(v=0)$ is not recommended since stimulated emission into the 4f hole states will reduce the photon scattering rate. Repumping via [31.05] is also a poor option, even though it connects strongly to both the 4f hole and X states, because this will transfer population to $\text{4f}_{7/2,3/2}^{-1}$. A better option is to repump via [18.58] and/or [18.71]. Our work establishes the required laser frequencies for this repump scheme, which are listed in Tab.~\ref{tab:repumps}. Closing these leaks will require either four separate lasers, or two lasers with high frequency modulators, alongside the four lasers used in earlier work to address $\text{X}^{2}\Sigma^+(v=0,1,2,3)$~\cite{Lim2018,Alauze2021}. Based on previous measurements and calculations of branching ratios~\cite{Zhuang2011,Smallman2014,Zhang2022}, it should then be possible to scatter about $10^{5}$ photons, which is sufficient for radiation pressure slowing and magneto-optical trapping.

As well as facilitating experiments in the ground state of YbF, it is interesting to consider whether the low-lying 4f hole states may themselves be useful for tests of fundamental physics. Electron EDM experiments measure the interaction energy between the EDM and an effective electric field, $E_{\rm eff}$, which is enormous in heavy polar molecules. When the molecule has a pair of near-degenerate states of opposite parity, a small applied electric field fully polarizes the molecule producing a pair of states where $E_{\rm eff}$ is equal and opposite. This ability to reverse the sign of the interaction through the choice of state is exploited in the most precise electron EDM measurements~\cite{Andreev2018,Roussy2023}. The ground states of laser coolable molecules do not have this structure, so attention has recently turned to polyatomic molecules~\cite{Hutzler2020}. The present work shows that the 4f hole states {\it do} have this structure providing long-lived states with closely-spaced parity doublets in a laser coolable molecule. The sensitivity to the EDM is proportional to $E_{\rm eff}$, but this can only be large when relativistic effects are strong, as happens for electrons with low orbital angular momentum ($l$) in heavy atoms and molecules. In a single electron picture, $E_{\rm eff}$ is proportional to $\bra{n l}H_{P,T}\ket{n' l'}$ where $H_{P,T}$ is the P,T-violating interaction, $l'=l\pm 1$ and only the small part of the Dirac wavefunction is involved~\cite{Kozlov1995}. It is usually $\bra{n \text{s}}H_{P,T}\ket{n' \text{p}}$ that contributes to make $E_{\rm eff}$ large because these electrons have substantial amplitude close to the nucleus where the electric field is large and relativistic effects are strongest. For the 4f hole states, $E_{\rm eff}$ is proportional to $\bra{\text{4f}}H_{P,T}\ket{n'\text{d}}$ which would appear to be heavily suppressed since these states have small amplitude near the nucleus. However, similar arguments can be made for the spin-orbit interaction which is also a relativistic effect and also decreases rapidly with $l$, and yet the spin-orbit splitting between the $\text{4f}^{-1}_{5/2}$ and $\text{4f}^{-1}_{7/2}$ states of YbF is very large, about 10,100~cm$^{-1}$. This is considerably larger than, for example, the spin-orbit splitting of the 5d and 6p states of Yb$^{+}$, which are 1372 and 3330~cm$^{-1}$ respectively. This suggests that the 4f electron sees a large effective nuclear charge because of its lower principal quantum number, becoming highly relativistic despite the larger angular momentum barrier that prevents its close approach to the nucleus. The same effect may also make $E_{\rm eff}$ larger than expected. Strong correlation effects within the 4f shell and between the 4f and outer electrons might also substantially alter the result compared to the single-particle picture. Thus, we think there is a strong case for calculating $E_{\rm eff}$ for this system. Furthermore, the 4f hole states may prove to be useful sensors of electric and magnetic fields, which is important for precision measurement. Closely-spaced parity doublets produce linear Stark shifts at low electric fields, and some of the 4f hole states should have strong Zeeman shifts. Measurements of these Stark and Zeeman shifts would be valuable. Finally, we note the considerable interest in developing molecular clocks, both as frequency standards and for measuring variations of fundamental constants~\cite{Kondov2019, Barontini2022}. The transition from the ground state to one of the 4f hole states may be suitable for such a clock, similar to the Yb$^+$ clock based on the equivalent electric octupole transition~\cite{Huntemann2012}. A measurement of the lifetimes of the $\text{4f}^{-1}_{7/2}$ states would be valuable.

[31.05] and [48.73] have a very similar rotational structure to the low-lying $\text{4f}^{-1}$ states. It is likely that they also correspond to excitations out of the 4f shell, though we do not currently know their underlying configurations. [48.73] is especially notable as it conforms almost exactly to the Hund's case (c) model with $|\Omega|=1/2$ and $J_a=3/2$. A broad-band spectroscopic survey of the entire set of inner-shell excited states in YbF would be useful, as would quantum chemistry calculations of this set. Through this combination of spectroscopy and calculation, it should be possible to determine the character of each state. 

We have found many electronic states above the ionization energy, including [48.72] and [48.73] studied here. Some of them have remarkably long lifetimes, and we suppose that these also correspond to excitation of a 4f electron. We attribute the existence of such strong and narrow auto-ionizing resonances to the weak coupling between the two series of electronically excited states. [48.73] has a 4f hole but does not have enough energy to auto-ionize to any 4f hole state of the YbF$^+$ cation or to dissociate to a state that has a 4f excitation in the atom or ion. Thus, [48.73] can only auto-ionize into the closed-shell ground-state of YbF$^+$, which requires a two-electron Auger process. Indeed, the first core-hole excited state of the cation can be considered as the effective ionization potential of the series of YbF 4f hole states, and is likely to lie approximately 8000-10000~cm$^{-1}$ above the normal ionization potential of the molecule. We advocate further studies of this interesting series of auto-ionizing states. The existence of strong auto-ionising states may open a channel for efficient state-selective detection of YbF by CW laser excitation to an auto-ionizing state. Precision measurements using this molecule currently use fluorescence detection which is often limited to an efficiency of about 10\%~\cite{Ho2020}, whereas ion detection can have almost unit efficiency if there is sufficient laser power for ionization. Finally, we note that these resonances result in an ionization spectrum that is different for each rotational state, which can be helpful in assigning quantum numbers to a spectrum. This method proved to be an extremely useful tool for elucidating the energy level structure of the 4f hole states from their complicated spectra.

\begin{acknowledgments}

We are grateful to Ben Sauer, Leonid Skripnikov and Lan Cheng for helpful discussions and advice. We acknowledge the assistance of Luca Diaconescu in some of the measurements. This research has been funded in part by the Gordon and Betty Moore Foundation through Grants 8864 \& GBMF12327 (DOI 10.37807/GBMF12327), and by the Alfred P. Sloan Foundation under Grants G-2019-12505 \& G-2023-21035, and by UKRI under grants EP/X030180/1 and ST/V00428X/1. 

\end{acknowledgments}

\bibliography{references}

\end{document}